\documentclass{PoS}
\usepackage{epsfig}

\newcommand{\kms}{\mbox{\rm km\,s$^{-1}$}}

\newcommand{\HI}{\mbox{H\,{\sc i}}}

\newcommand{\am}[2]{$#1'\,\hspace{-1.7mm}.\hspace{.0mm}#2$}

\def\lsim{~\rlap{$<$}{\lower 1.0ex\hbox{$\sim$}}}
\def\gsim{~\rlap{$>$}{\lower 1.0ex\hbox{$\sim$}}}

\PoS{PoS(BDMH2004)070}
\title{Clues on the Structure and Composition of Galactic Disks from 
Studies of `Superthin' Spirals: the Case of UGC~3697}
\ShortTitle{UGC~3697}
\author{\speaker{Lynn D. Matthews}\\
        Harvard-Smithsonian Center for Astrophysics, USA\\
        E-mail: \email{lmatthew@cfa.harvard.edu}}
\author{Juan M. Uson\\
         National Radio Astronomy Observatory, USA\\ 
        E-mail: \email{juson@nrao.edu}}

\abstract{We summarize results from an \HI+optical imaging study of
the ``Integral Sign'' galaxy, UGC~3697. UGC~3697 is a low-mass, Sd
spiral that exhibits a ``superthin'' disk morphology despite a
prounced gasous and stellar warp. Our new observations show
evidence for a recent minor merger in this system that could account
for its large-scale warp and a number of other properties of this
galaxy. We speculate that UGC~3697 has been caught in a rather short-lived
dynamical state, and may soon undergo signi${\rm f}$icant structural and
morphological 
changes.}

\FullConference{Baryons in Dark Matter Halos\\
 5-9 October 2004\\
 Novigrad, Croatia}

\begin{document}
\begin{center}
\epsfysize=5.0 cm
\epsfbox{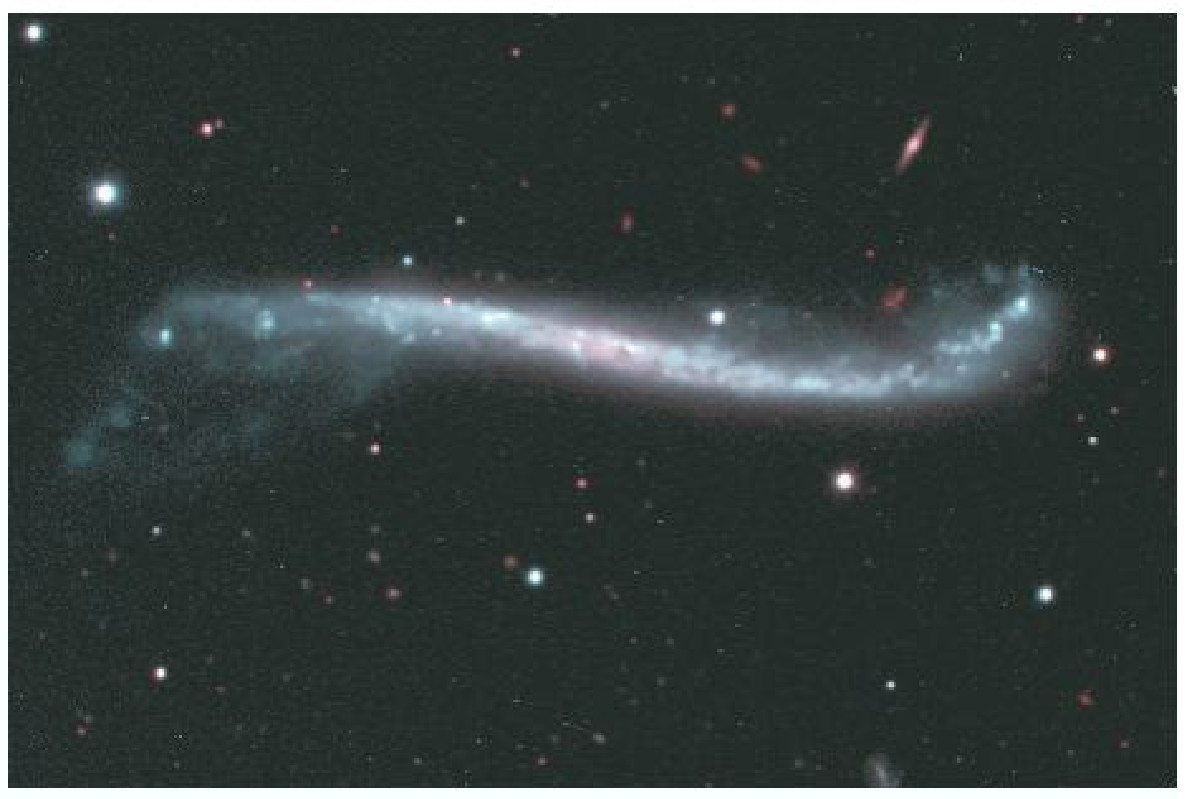}
\end{center}
\vspace{-0.5cm}
{\small\em {\bf Figure 1}. Optical $B+R$ composite image of UGC~3697
obtained with the 3.5-m WIYN telescope at Kitt Peak, AZ. The $f$ield
shown is $\sim$\am{3}{5} across.}

\section{Background}
Among the latest-type spiral galaxies, there exist systems that when
viewed edge-on, exhibit extraordinarily thin stellar disks
($h_{z}\sim$150-200~pc; $a/b>10$), often of very low
intrinsic surface brightness ($\mu(0)_{B,i}\gsim$23 mag arcsec$^{-2}$). 
The chemical, structural, and dynamical properties of
these so-called ``superthin'' galaxies imply
they are some of the least evolved disk galaxies in the local
universe, both in terms of 
their dynamical properties and their star formation histories 
(\cite{goad},\cite{bergvall},\cite{matthews99}). 
This implies that the superthins 
can supply important constraints on the process of
galaxy disk formation, the role of environment on the evolution of disk
galaxies, and on the nature and distribution of galactic dark
matter (e.g., \cite{karachentsev},\cite 
{matthews99},\cite{uson}).
Recently, we have been investigating these issues using a combination of
optical and \HI\ imaging  of
three nearby superthin galaxies
(\cite{uson},\cite{matthews05},\cite{mu}). Here we
describe preliminary results from the analysis of one of these
systems, UGC~3697.

UGC~3697 has been referred to as the ``Integral Sign'' galaxy
(\cite{burbidge}) owing to the pronounced $S$-shaped bending of its
stellar disk (Fig.~1). Although some degree of warping is 
observed in the stellar and/or gaseous  disks of a large fraction of 
galaxies (\cite{reshetnikov99}), few are are warped
as dramatically  as 
UGC~3697. The rarity of warps with similar amplitudes ($\sim$15\%
of the disk diameter)
suggests UGC~3697 is in a transient dynamical state.

Despite the ubiquity of galactic warps, their origin is still poorly
understood (\cite{binney}). Accretion events are one means 
of exciting high-amplitude warps (e.g., \cite{diaz}); 
however, accretion and minor mergers are expected
to  heat (thicken) disks dynamically 
(e.g., \cite{walker}), leading to the puzzle of how
a galaxy like 
UGC~3797 can be so strongly warped, while preserving such a thin disk
structure. 

Table~1 presents a comparison between some of the global properties of
UGC~3697 and a prototypical superthin galaxy, UGC~7321
(\cite{matthews99}, \cite{uson}). Note that both galaxies have
extremely small stellar scale heights, and share a number of other 
physical similarities
(e.g., linear diameter; peak rotational velocity; \HI\ mass).
Both UGC~7321 and UGC~3697 are also pure disk systems with no
obvious equatorial dust lane. On the other hand, UGC~3697 is
roughly three times more luminous than UGC~3697 
in the optical, and 18 times more
luminous in the FIR. Its central CO luminosity is signi{\rm f}icantly higher
(\cite{mu}), and its inner rotation curve
much steeper (\cite{goad}). 
The combination of these similarities and differences strongly suggests that
UGC~3697 is a typical superthin system in the throes of a structural
and morphological
transformation. 

\section{Results from New \HI\ and Optical Imaging}
Fig.~2 shows an \HI\ total intensity image of UGC~3697 obtained with the
Very Large Array (VLA).\footnote{The Very Large Array
of the National Radio Astronomy Observatory is a facility of the
National Science Foundation, operated under cooperative agreement by
Associated Universities, Inc.} 
These data were obtained in the C con${\rm f}$iguration
with 12 hours of on-source integration, and reach a 3$\sigma$
limiting column density of $\sim2\times10^{19}$ atoms cm$^{-2}$
channel$^{-1}$ with a velocity resolution of 5.2~\kms\ and a spatial
resolution of $\sim19''$.

\begin{table*}
\begin{scriptsize}
%\addtocounter{table}{1}
%\vspace{-10.0cm}
\begin{center}
\centerline{Table 1.}
\begin{tabular}{|lll|}
\hline
\multicolumn{1}{|l}{} &
\multicolumn{1}{l}{UGC~3697}    & \multicolumn{1}{l|}{UGC~7321}   \\
\hline
$D$(Mpc) & 18 & 10  \\
$V_{\rm rot}$(\kms) & 95 & 104  \\
$A_{25}$(kpc)& 17.0 & 16.1  \\
$h_{z}$(pc) & 200 & 150  \\
$M_{\rm HI}$($\times10^{9}M_{\odot}$) &   1.4 & 1.1  \\
$L_{B}$($\times10^{9}L_{\odot}$) & 2.8 & 1.0  \\
$L_{FIR}$($\times10^{9}L_{\odot}$) &  1.2 & 0.070  \\

\hline
\end{tabular}
\end{center}
\vspace{-1.0cm}
\end{scriptsize}
\end{table*}

Fig.~2 reveals that \HI\ traces the stellar warp of UGC~3697, and also shows
additional twists and
extensions. On the eastern side of the galaxy, we {\rm f}ind gas
along the midplane, as well as a wide swath of more
diffuse emission sweeping below the plane. In our optical images
(Fig.~1), this latter 
region is faintly delineated by a very blue sheet of stars. Near
this location, we also  draw
attention to a faint, newly-discovered  dwarf
near the southeastern edge of the disk, at ($\alpha_{2000}$=07$^{\rm
h}$11$^{\rm m}$57.1$^{\rm s}$,
$\delta_{2000}$=71$^{\circ}$48$^{'}$55$^{''}$). 
This dwarf shows
a rotational signature, and has an optical 
counterpart $\sim10''$ across,
with $L_{B}\sim8\times10^{6}L_{\odot}$.

In contrast to 
normal edge-on spirals, where the brightest concentration of
\HI\ is typically found near the central regions of the galaxy, UGC~3697
shows a bright blob of emission near its western
edge. Gas {\rm f}ilaments also extend from this
location, reaching up to $\sim$7~kpc from the plane.  Neither the
{\rm f}ilaments nor the bright midplane clump have any obvious optical
counterparts.

\vspace{-0.5cm}

\section{Interpretation}

%\vspace{-1.0cm}

While UGC~7321 is a rather isolated system (\cite{uson}), 
UGC~3697 is part of a small galaxy group that includes
the peculiar elliptical UGC~3714 and several dwarf galaxies
(\cite{garcia}). UGC~3714, at a projected distance of 39~kpc from UGC~3697, is
the optically brightest group member, and has long been presumed to be
responsible for exciting the warp of UGC~3697 via tidal effects
(e.g., \cite{nilson}). However, tides are rather ine{\rm ff}icient
warp drivers (e.g., \cite{garciaruiz}), 
and our new observations have for the {\rm f}irst time
revealed strong
evidence that an accretion event rather than tidal effects are
responsible for the present morphology of UGC~3697.
 
We estimate that a mass of \HI\ of a few times
$10^{8}M_{\odot}$ has been recently accreted onto
UGC~3697 from an infalling satellite. This mass is typical of the \HI\
contents of dwarf irregular galaxies. Such an
event would also be expected to trigger radial gas in${\rm f}$lows
(\cite{hernquist}) and may thus account for the centrally-enhanced CO and 
21-cm radio continuum
emission (\cite{mu}), 
as well as the steep appearance of its inner rotation curve
compared with other superthins
(see \cite{diaz}).  
The faint dwarf we see near the southeastern tip
of UGC~3697 may be the stripped core of this intruder. 

While the infall of satellites is expected to 
heat (thicken) disks dynamically, the simulations of 
\cite{sellwood} showed
that disks can remain thin during several passes of a low-mass intruder,
before rapidly thickening via
resonantly-excited bending waves once the satellite 
orbit has decayed  suf{\rm f}iciently. 
We speculate that UGC~3697 may be caught in a
rather special state, en route to such a
trasnformation. Such events are likely to have been common in low-mass
disk galaxies during earlier epochs, implying studies of galaxies like
UGC~3697 can shed insight into the evolutionary consequences of this process.
Moreover, because superthin systems like UGC~3697 are structurally simple
and 
dark matter-dominated, further observations and modelling of 
galaxies like these 
may supply important constraints on the amount of dark
matter that resides in the disk versus the halo of late-type spirals. 

\begin{center}
\epsfysize=5.5cm
\epsfbox{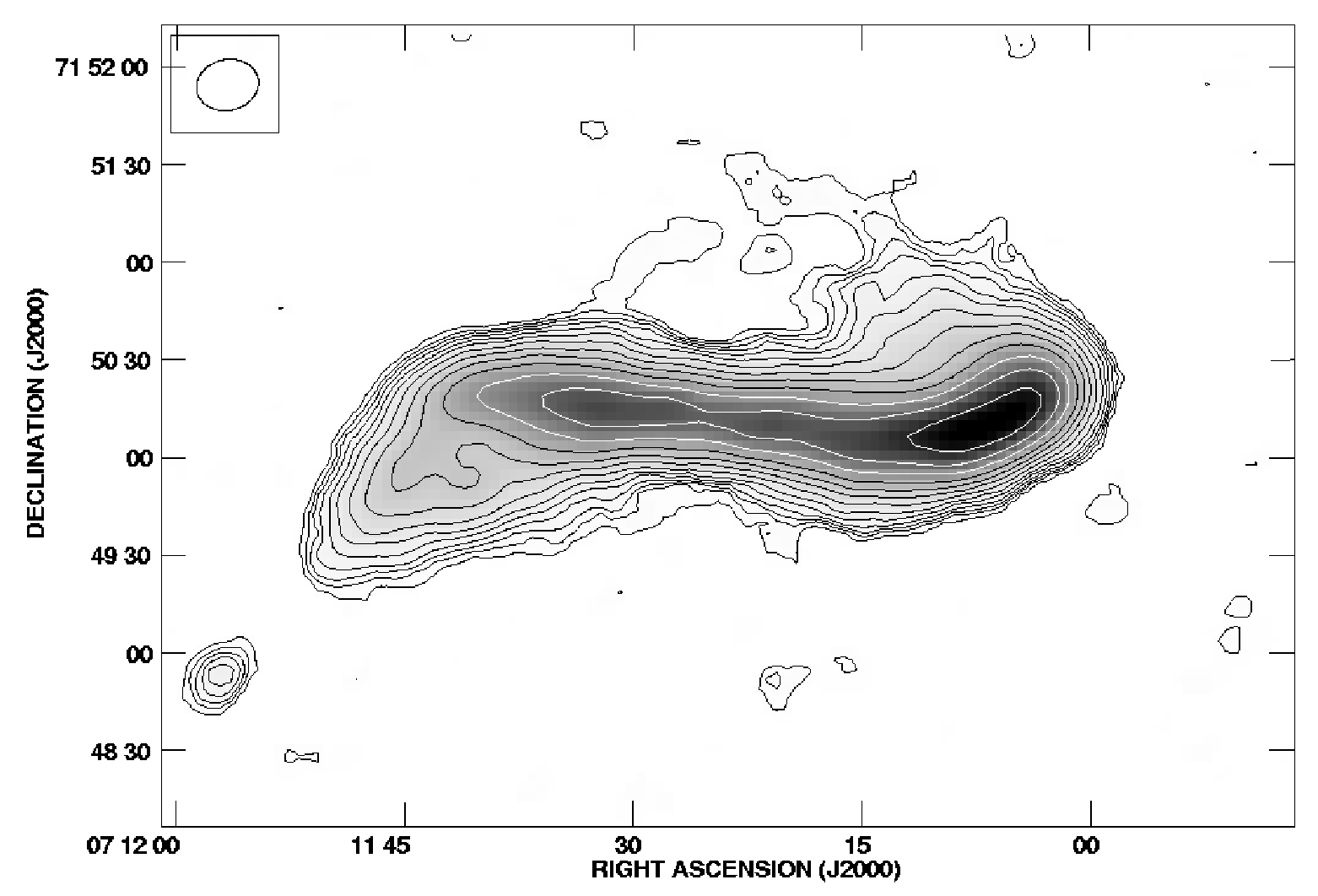}
\end{center}
\vspace{-0.5cm}
{\small\em {\bf Figure 2} Contour and greyscale representations of
the \HI\ intensity distribution in UGC~3697 based on VLA data. A
previously uncatalogued dwarf is visible in the lower left. The
contour levels are:
%0.52,1.0,1.5,2.1,2.9,4.0,5.9,8.5,12,17,24,33,47,\& 67$\times10^{20}$
0.52,1.0,1.5,2.1,2.9...67 $\times10^{20}$
cm$^{-2}$ at a resolution of $19''\times 16''$
(1.7$\times$1.4~kpc). The greyscale range is 0-5.5$\times10^{21}$ cm$^{-2}$.}

\end{document}